# The Great Wall: Urca Cooling Layers in the Accreted NS Crust

Zach Meisel[1,*]

[1]Department of Physics and Astronomy, Ohio University, 45701 Athens, Ohio USA

**Abstract.** Accreting neutron stars host a number of astronomical observables which can be used to infer the properties of the underlying dense matter. These observables are sensitive to the heating and cooling processes taking place in the accreted neutron star (NS) crust. Within the past few years it has become apparent that electron-capture/beta-decay (urca) cycles can operate within the NS crust at high temperatures. Layers of nuclei undergoing urca cycling can create a thermal barrier, or Great Wall, between heating occurring deep in the crust and the regions above the urca layers. This paper briefly reviews the urca process and the implications for observables from accreting neutron stars.

## 1 Introduction

Describing how matter behaves across the range of temperatures and densities occurring in nature is the ultimate goal of much of modern physics research. The only avenue presently available to study matter at the highest densities, near and above those of an atomic nucleus, at relatively low temperatures is the study of neutron stars. Neutron stars accreting material from a binary companion are particularly useful in this regard, as their observable phenomena provide clues as to the underlying structure of the neutron star (NS).

Gleaning information from such observables requires comparisons between the observed data and results from astrophysics model calculations. However, successful reproduction of the observed data requires high quality physics input. As such, the model-observation comparison process is inherently iterative, where the model input physics is adjusted until agreement is achieved. Unfortunately, many degeneracies exist and so efforts must be made to reduce the number of free parameters in models by constraining the physics input as best as possible.

While much of the physics of dense matter is beyond the reach of terrestrial experiments, and so only theoretical constraints can be obtained, the nuclear physics of relevance for the NS outer layers can often be measured. In particular, the nuclear reactions of the outer crust and above can be constrained using indirect measurements at present stable and radioactive ion-beam facilities.

The nuclear physics of particular interest for this work is the phenomenon of electron-capture/$\beta^-$-decay cycling, known as urca cooling, which was recently found to be active in the crusts of accreting neutron stars. Prior to describing urca cooling and its impact, a brief discussion of reactions in the outer NS (shown schematically in Figure 1) will help provide context.

## 2 Journey of a Nucleus in the Outer NS

Low mass X-ray binary systems which host observables such as type-I X-ray bursts, X-ray superbursts, and decaying X-ray emission from NS crust cooling, are generally thought to involve accretion of hydrogen and/or helium-rich material onto the NS surface. The H/He fuel builds-up on the NS surface, bringing in roughly 200 MeV per accreted nucleon. The combination of heat and fuel results in nuclear burning on and near the NS surface which produces a variety of heavier nuclei. A number of burning processes produce iron-rich ashes, though a wide variety of nuclei up to the tin region can be made, including special burning regimes which enrich the NS ocean in carbon [1–4].

Subsequent accretion buries the surface burning ashes to deeper, denser conditions. The electron Fermi energy $E_{e^-,Fermi}$ rises with increasing depth, so that buried nuclei eventually encounter a region where the degenerate electron gas can pay the energy debt required for electron capture $Q_{EC}$. Near $E_{e^-,Fermi} \approx Q_{EC}$, electron-capture EC proceeds, transforming a proton within a nucleus into neutron. Therefore, in the upper regions of the outer crust, where EC is the dominant reaction mechanism, buried ashes maintain their original abundance distribution in terms of mass number A, but are transformed into more and more neutron rich isobars.

Eventually, EC has driven the composition near to the neutron-drip line, so that neutron emission via (EC,$x$n), where $x$ is some integer, can occur. The neutron-drip point marks the transition between the outer and inner crust. At similar and deeper densities, the zero-point vibrations of nuclei are sufficient to tunnel through the energy barrier between them, so that density driven (pycnonuclear) fusion proceeds. Ultimately the nuclei that were born on and near the NS surface will dissolve into the core once buried to a depth near nuclear saturation density. The entire journey takes on the order of a million years for an NS continually accreting near the Eddington accretion rate (The rate at which the force of radiation emitted from accretion is balanced with the inward force of gravity).

---

[*] Email: meisel@ohio.edu

The combination of EC and fusion reactions just described dramatically transforms the outer NS composition and drives it from thermal equilibrium [5].

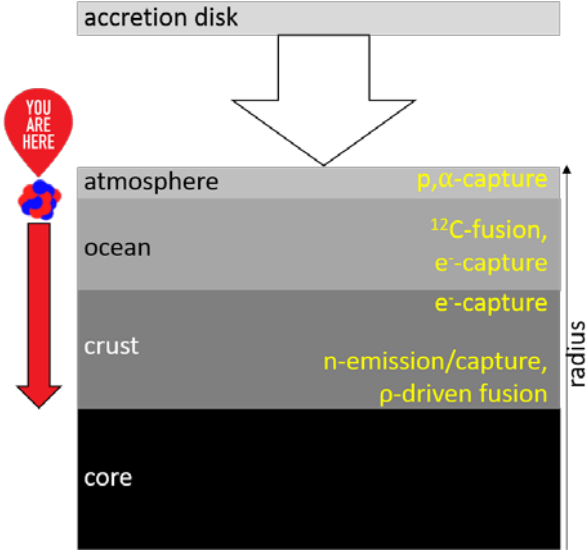

**Fig. 1.** Nuclear processes relevant for a nucleus as it is buried by accretion to various depths in the NS outer layers.

## 3 Urca Cycling and the Great Wall

For most of the time since EC reactions in the NS crust were hypothesized, they were considered to be heat sources. In particular, EC reactions to excited states, which are likely prevalent in two-step EC reactions that occur for even-A isobars, can release several-hundred keV per EC [6]. The reverse process of β⁻-decay went mostly ignored since the electron-degenerate environment implies most of the phase-space is blocked for an electron to re-enter the environment. However, recent calculations of the thermal and compositional evolution of an accreting crust employing a multi-species nuclear reaction network showed that β⁻-decay is likely possible for a large number of nuclei [7]. Once β⁻-decay proceeds for such cases, EC is still energetically favorable and so the process can repeat cyclically. This is generally limited to odd-A nuclides.

EC/β⁻-decay cycling was proposed for dense stellar interiors by Gamow and Schoenberg, who coined it the urca process [8,9]. The cycling occurs as shown in Figure 2, where EC proceeds near $Q_{EC}$ to the ground or a low-lying excited state of the EC daughter. Low-lying states of the EC parent can also participate due to thermal population. In each case, "low-lying" means on the order of $k_BT$, where $k_B$ is the Boltzmann constant and T is the environment temperature. The neutrino luminosity $L_\nu$ for isotopes undergoing urca cycling, referred to as an urca pair, is proportional to the weak transition rate between the nuclides, as quantified by the comparative half-life $ft$; the phase-space for the weak transition, which is sensitive to $Q_{EC}$; and the thickness of the urca layer, which is proportional to $k_BT$ and inversely proportional to the local acceleration due to gravity $g$.

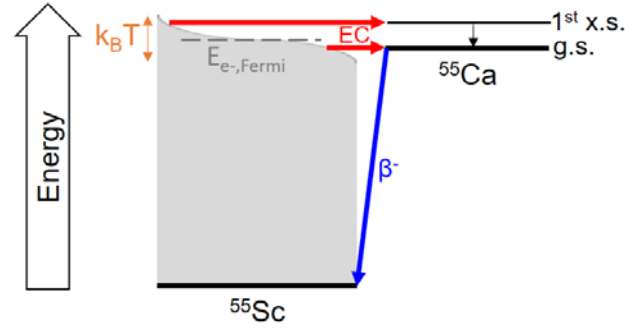

**Fig. 2.** Urca cycling between $^{55}$Sc and $^{55}$Ca at the depth in the NS crust where $E_{e^-,Fermi} = Q_{EC}(^{55}Sc)$, including a hypothetical low-lying excited state in $^{55}$Ca which is only included for illustrative purposes.

Quantitatively, for an EC parent $^AZ$ [10,11]: $L_\nu(Z,A) \approx L_{34}(Z,A)X(A)T_9^5(2/g_{14})(R/10km)^2 \cdot 10^{34}$ erg/s, where X(A) is the mass fraction of A, $T_9$ is the local temperature in gigakelvin, $g_{14}$ is $g/10^{14}$ cm/s², and R is the NS radius in kilometers. $L_{34}(Z,A)$ is the intrinsic cooling strength for an urca pair with EC parent Z,A: $L_{34} = 0.87(10^6 s/ft)(56/A)(Q_{EC}(Z,A)/4MeV)^5(2<F>^*)$, where $<F>^* = (<F>^+<F>^-)/(<F>^+ + <F>^-)$, the Coulomb factor $<F>^\pm \approx 2\pi\alpha/|1-exp(\mp 2\pi\alpha Z)|$, and α is the fine structure constant. For this equation $ft$ is the average $ft$-value for the EC and β⁻-decay, which can differ slightly due to different state degeneracies.

For context, typical crustal heating is, depending on the heat deposited per accreted nucleon and the accretion rate, on the order of a few ×10³⁴–10³⁶ erg/s [7]. In practice, urca cooling impacts the local environment temperature significantly enough to impact observables (see the following section) when the temperature is around ~1GK and $X(A)L_{34}(Z,A) \gtrsim 1$. For example, if X(A) ~ 1%, this would be roughly satisfied if $Q_{EC}$ ~ 8 MeV and log($ft$) ~ 5, or if $Q_{EC}$ ~ 15 MeV and log($ft$) ~ 6.5. Of course, combinations of larger $Q_{EC}$ and smaller log($ft$) result in larger $L_\nu$.

Estimates of $L_\nu$ are sensitive to nuclear physics input through $Q_{EC}$, $ft$, and X(A). For the majority of nuclei in the NS crust, $Q_{EC}$ is known to a few percent, though theoretical predictions for unmeasured cases can differ by more than 10%. $Ft$, however, is generally unknown and is often uncertain by several orders of magnitude. X(A) relies on results from model calculations of surface nuclear burning processes, which have many astrophysics uncertainties, but even the uncertainties in nuclear reaction rates can impact X(A) by orders of magnitude [12].

As an example, consider the urca pair $^{55}$Sc–$^{55}$Ca. The atomic mass excess ME of $^{55}$Sc is known to roughly half an MeV [13] and the mass of $^{55}$Ca is unmeasured, though the disagreement between theoretical predictions in this region can be on the order of an MeV. As such, the uncertainty for $Q_{EC}$ ( = ME(Z,A) - ME(Z-1,A)) in this case is roughly 13%, resulting in around a factor of 2 uncertainty in $L_\nu$. *Ft* has not been measured for the ground-state to ground-state transition between $^{55}$Sc and $^{55}$Ca, so an estimate must be made. QRPA and shell-model calculations for this transition predict log(*ft*) = 4.2 and 3.9, respectively [7,14]. Alternatively, a data-based systematics approach [15] can be used (which avoids costly, large-scale calculations). From a compilation of log(*ft*) for transitions with a known change in spin *J* and parity π, log(*ft*) can be adopted based on a transition's $\Delta J^{\Delta\pi}$. For the $^{55}$Sc–$^{55}$Ca ground-state to ground-state transition, $\Delta J^{\Delta\pi}$ = 1$^+$, according to tentative $J^\pi$ assignments [16]. From a compilation [17], it is apparent that the range of log(*ft*) for $\Delta J^{\Delta\pi}$ = 1$^+$ for ~700 odd-A nuclei is between 3.5–9.1, where the distribution is centered about log(*ft*) = 5.9 ± 1. Thus, a conservative estimate for the uncertainty contribution of *ft* to $L_\nu$ is a factor of 10. Estimating the uncertainty in X(A) is very difficult, though an empirical upper-limit for odd-A nuclides appears to be somewhere near X(A) < 10%. For X-ray bursts it is likely more than an order of magnitude due to nuclear reaction rate uncertainties alone [12]. For X-ray superbursts, the X(A) distribution is mostly determined by nuclear statistical equilibrium [18], and hence relatively well known ME for nuclei around stable iron, but the uncertainty contribution to late-time H/He capture reactions on superburst ashes has not been explored. Therefore, it is clear that a number of nuclear physics experiments are needed to adequately constrain $L_\nu$ for this case and many more like it (see section 5).

The consequence of including an urca layer in the NS crust is that a high-temperature thermostat exists at that depth. The $T^5$ dependence ensures that any significant heat flux encountering the urca layer will be bled away by neutrinos. This amounts to a thermal barrier, or Great Wall, between the deep crust, where (EC,*xn*) and pycnonuclear fusion occur, and the ocean and atmosphere where H/He and C burning drive bursts. As an example, Figure 3 shows the impact of an urca layer for an NS crust heated to over 1 GK by accretion.

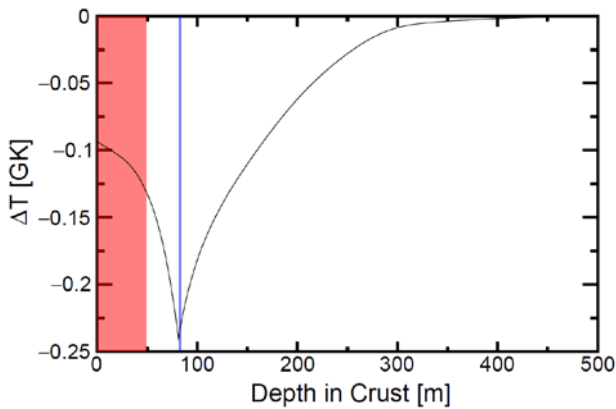

**Fig. 3.** Residual for the NS crust temperature after 480 days of accretion at 1.3×10$^{18}$ g/s, assuming a shallow heating (red region) of 8 MeV per accreted nucleon in a crust with impurity parameter Q$_{imp}$=1, between `dStar` [19] models with and without Urca cooling. The cooler model adopted an urca layer (blue line) mimicking $^{55}$Sc–$^{55}$Ca cycling assuming log(*ft*)=4.9 and X(55) =1.8%.

# 4 Implications for Observables

A number of accreting NS observables are sensitive to the thermal properties of the NS outer layers. The ignition of X-ray bursts and superbursts are sensitive to the heat flux from the underlying NS, often referred to as base heating. Cooling transients directly probe the NS composition and temperature as a function of depth, since the X-ray flux at any given point in the cooling light curve originates from the depth with which the surface has reached thermal equilibrium. It is then clear that urca cooling is of interest for accreting NS model-observation comparisons.

For X-ray superbursts, an unphysical urca cooling $L_\nu$ would be required to alter the light curve. However, X(A)L$_{34}$(Z,A) $\gtrsim$ 10 is sufficient to modify the depth of the carbon ignition which powers superbursts. Within estimated uncertainties for log(*ft*), the superburst ignition depth predicted from model calculations could be lowered to more than an order of magnitude larger pressure [11]. Interestingly, the ignition depth inferred from observations is already an order of magnitude shallower than found in models [20]. An as-yet unidentified shallow heating source has been invoked to reconcile the discrepancy (and is also required for adequate reproduction of most cooling transients [21] and some type-I X-ray bursters [22]). As A=55 ashes are robustly produced in calculations of superburst ashes, present nuclear physics estimates would require the shallow heat source be located above the $^{55}$Sc EC depth, as deeper heat sources would be thermally isolated from heating the NS ocean due to $^{55}$Sc–$^{55}$Ca urca cooling.

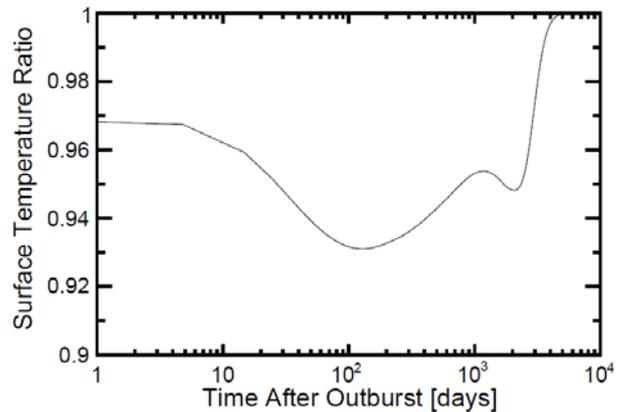

**Fig. 4.** Ratio between the surface temperatures for the `dStar` models described in Figure 3 for the time after accretion has ceased. The hundred-day dip is directly due to the inclusion of the urca layer, whereas the feature near one thousand days is due to the change in absolute time when the surface reaches equilibrium with deep crustal heating when an urca layer is included. Note that typical observational uncertainties are on the percent level [23,24].

The impact on cooling transients is limited to NS crusts which have been heated out of equilibrium to a temperature of ~1 GK, as is thought to be the case for the source MAXI J0556-332 for its earliest observed crust cooling phase [23,24]. For such a hot NS crust, an urca layer with X(A)L$_{34}$(Z,A) $\gtrsim$ 1 is sufficient to appreciably alter the outer NS crust temperature profile [15] (Figure

3 shows an example with X(A)L$_{34}$(Z,A) ≈ 44). Due to the associated modifications in thermal gradients in the NS crust, heat transport time-scales are modified and therefore the rate at which the surface reaches thermal equilibrium with particular depths is changed. The result is a dip that is introduced in the crust cooling light curve around ~100 days after accretion ceases (see Figure 4). Though a large number of poorly constrained astrophysics inputs contribute to uncertainties in models of cooling transient light curves [21], and many inputs are degenerate in their impacts, no other known mechanism exists to mimic the hundred-day dip. Therefore, a positive identification of such a dip would be direct evidence for urca cooling in the crust, possibly validating surface nucleosynthesis models.

Though urca cooling has gone undetected, one can use the negative result for MAXI J0556-332 to place interesting constraints on this source. As the urca mechanism is robust when using a wide range of nuclear physics and astrophysics model assumptions, the lack of an urca cooling signature in the cooling transient light curve for an NS with a hot crust implies that urca nuclei are not present at $E_{e-,Fermi} \approx Q_{EC}$. This then implies that these nuclei were not produced on or near the NS surface in the past, where the exact time constraint depends on the assumed past accretion rate and duty cycle. Since both X-ray burst and superburst ashes are thought to produce urca nuclei with significant associated $L_v$, we can conclude that MAXI J0556-332 did not exhibit type-I X-ray bursts or superbursts in the past centuries to millenia (or at least not for sustained periods) [15]. This source presently exhibits stable burning near the Eddington accretion rate, which is not thought to yield significant abundances of urca nuclides [1,15]. Taken with the previous conclusion, this implies MAXI J0556-332 has not reduced its accretion rate to the roughly tenth of Eddington rate needed to have bursting and supersursting behaviour.

Since high accretion rates are needed to reach gigakelvin crust temperatures and stable burning proceeds at these rates [1], it becomes clear that a special sort of NS source would be required to result in a detectable urca signature in its cooling transient light curve. The source must accrete at a low enough rate to exhibit bursts and/or superbursts for centuries so that a $~k_BT$-wide urca layer can be built-up, then accrete at a near-Eddington accretion rate for at least a year in order to heat the crust to ~gigakelvin temperatures, and then cease accretion for several hundred days or more so that the crust cooling phase can be observed. Though seemingly contrived, a source that appears to almost fulfill this criterion was already observed, though KS1731-260 appears to host relatively modest shallow heating [25,26] and therefore too low of an NS crust temperature to exhibit significant urca cooling luminosities. The observation of cooling transients is less than two decades old, with most model-observation comparisons taking place in the last decade, so the existence of a source fulfilling the previously stipulated conditions is a distinct possibility.

## 5 Directions for Future Research

As demonstrated, urca cooling has significant implications for our interpretation of observables from accreting neutron stars. Positive identification of an urca signature would provide unique evidence of nuclear reactions in the NS crust, whereas the lack of an urca signature where it is expected would require a change in our picture of the accreting NS outer layers. However, observational data for phenomena sensitive to urca cooling are limited, systematic studies of degeneracies between urca cooling signatures and other astrophysical effects have not been performed, and nuclear physics inputs that determine $L_v$ are mostly poorly constrained. This provides opportunities for continued work in experiment, theory, and observation.

The only observable phenomena known where a direct signature of urca cooling is expected is crust cooling from transiently accreting systems. However, less than ten such systems are known and only one appears to have had a crust hot enough for $L_v$ to be significant [27]. While this may be cause for concern, the research focus of cooling transient observations is in its relative infancy (e.g. compare to nearly half a century of type-I X-ray burst observations) and several groups are actively working to identify new objects and to monitor additional cooling events from known sources. Should another source as hot as the first observed outburst of MAXI J0556-332 be observed, it will be imperative for repeated observations to occur in the several tens to several hundreds of days after accretion ceases in order to definitively characterize the cooling light curve around the possible hundred-day dip.

The possibility of an urca signature in cooling transient light curves has only recently been identified and so few model calculations have investigated this phenomenon [15,28]. While it has been shown that an observable urca signature is possible, the range of signatures for different surface burning ashes and different accretion conditions has not been explored. Such systematics studies will be necessary to maximize the information obtained from model-observation comparisons should a transient be observed with the anticipated hundred-day dip in the light curve. Ideally, advances will be made in deducing surface burning conditions from model-observation comparisons of bursts on the NS surface. This would enable more reliable predictions for ash production on the surface of particular sources, and therefore remove degrees of freedom when modelling cooling transients. Indeed, such multi-observable modelling will likely be key to remove degeneracies that are present in the impact of astrophysics inputs on model calculation results.

The orders of magnitude uncertainties in $L_v$ from nuclear physics inputs (see section 3) will require a considerable experimental effort spanning much of the nuclear chart below A~100, as indicated by Figure 5.

Studies of surface burning processes, such as the rapid proton-capture (*rp-*) process that powers type-I X-ray bursts, must be sure to include investigations of odd-A nuclide production. While the majority of studies to date have focused on the impact of nuclear physics uncertainties on the X-ray burst light curve, this work shows that understanding odd-A nuclide production is just as critical. In particular, odd-A isobars that are suspected to have neutron-rich isotopes with relatively low log(*ft*) are of interest. These tend to be located in regions between nuclear closed shells, where there is appreciable deformation [7].

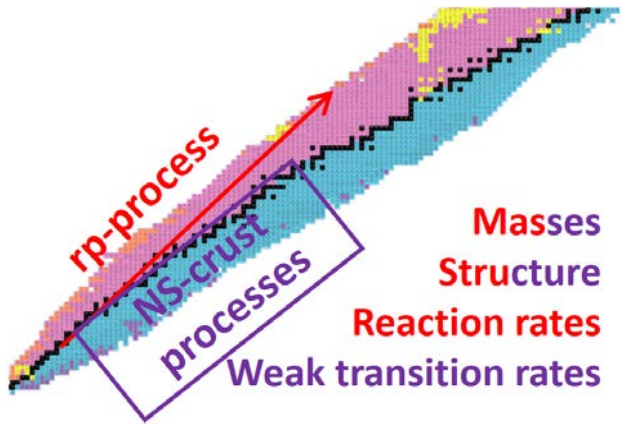

**Fig. 5.** Schematic indicating nuclear physics properties that need to be constrained to better estimate the $L_\nu$ associated with urca cooling in the NS crust.

Nuclear mass measurements are required to better constrain $Q_{EC}$, which impacts the strength and determines the location of urca layers. When low-lying excited states are suspected to play a role, higher precision masses ($\delta Q_{EC} \lesssim k_B T$) are necessary to determine the extent to which they can participate in urca cycling. This is also true for any even-A cases with anomalously small odd-even mass-staggering [13].

Weak transition rates, as quantified by *ft*, are arguably the most uncertain input in present estimates of $L_\nu$. Precise constraints will require $\beta^-$-decay measurements far off of stability which carefully account for all decay branches, including delayed neutron emitting channels. In their absence, spectroscopy measurements to determine $J^\pi$ of ground and low-lying states can be used to roughly estimate *ft* from known systematics related to $\Delta J^{\Delta \pi}$.

The above discussion leaves a long to-do list for nuclear physics experimentalists at both stable and radioactive ion beam facilities. Stable beam facilities will be able to map nuclear structure properties influencing nuclear reaction rates near the valley of stability, whereas radioactive ion beam facilities will be required to do most of the rest of the work. Of course, any improvements in nuclear theory would be welcome to provide best-guesses for any nuclear data that presently defies measurements.

These are exciting times for those interested in the physics of the accreting NS outer layers and the associated observables. The possible presence of Great Walls of urca cooling provides new challenges in experiment, theory and observation, which are sure to keep the field busy for some time.

## Acknowledgements

This work resulted from collaborations with Edward Brown, Andrew Cumming, Alex Deibel, and Hendrik Schatz. The author is supported in part by the U.S. Department of Energy through grant № DE-FG02-88ER40387. This material is based on work supported by the National Science Foundation through grant № PHY-1430152 (Joint Institute for Nuclear Astrophysics – Center for the Evolution of the Elements).